\documentclass[useAMS,usenatbib]{mn2e}
\usepackage{graphicx,natbib,color,multirow,amsmath,url, hyperref}
\usepackage{booktabs,caption,fixltx2e}
\usepackage[flushleft]{threeparttable}

\definecolor{titlecol}{rgb}{0,0,1}
\definecolor{titlecol2}{rgb}{0,0.65,0}
\definecolor{titlecol3}{rgb}{0.99,0.4,0.} 

\def\lesssim{\mathrel{\hbox{\rlap{\hbox{\lower3pt\hbox{$\sim$}}}\hbox{\raise2pt\hbox{$<$}}}}}
\def\gtrsim{\mathrel{\hbox{\rlap{\hbox{\lower3pt\hbox{$\sim$}}}\hbox{\raise2pt\hbox{$>$}}}}}

\newcommand{\ls}{{_<\atop^{\sim}}}
\newcommand{\gs}{{_>\atop^{\sim}}}
\def \spose#1{\hbox  to 0pt{#1\hss}}  
\def \ls{\mathrel{\spose{\lower 3pt\hbox{$\sim$}}\raise  2.0pt\hbox{$<$}}}
\def \gs{\mathrel{\spose{\lower  3pt\hbox{$\sim$}}\raise 2.0pt\hbox{$>$}}}

\newcommand{\OIII}{\hbox{[{\rm O}\kern 0.1em{\sc iii}]}}
\newcommand{\NII}{\hbox{[{\rm N}\kern 0.1em{\sc ii}]}}

-lmtk10

\newcommand{\apj}{ApJ}
\newcommand{\apjs}{ApJS}
\newcommand{\apjl}{ApJL}
\newcommand{\aap}{A{\&}A}

\newcommand{\mnras}{MNRAS}
\newcommand{\aj}{AJ}
\newcommand{\araa}{ARAA}
\newcommand{\pasp}{PASP}

\newcommand{\nat}{Nature}

\newcommand{\procspie}{Proc. SPIE}
\newcommand{\fcp}{Fund. Cosm. Phys}

\begin{document}

\title[Bar-AGN Connection at $0.2<z<1.0$]{Galaxy Zoo: Are Bars Responsible for the Feeding of Active Galactic Nuclei at $0.2<z<1.0$?\thanks{This publication has been made possible by the participation of more than 85,000 volunteers in the Galaxy Zoo project. Their contributions are individually acknowledged at http://authors.galaxyzoo.org/.} }

\author[Cheung et al.]{\parbox[t]{16cm}{Edmond Cheung$^{1,2}\thanks{E-mail: ec2250@gmail.com}$, 
Jonathan R. Trump$^{3,18}$, 
E. Athanassoula$^{4}$, 
Steven P. Bamford$^{5}$,
Eric F. Bell$^{6}$, 
A. Bosma$^{4}$,
Carolin N. Cardamone$^{7}$,
Kevin R. V. Casteels$^{8}$,
S. M. Faber$^{1,9}$,
Jerome J. Fang$^{1}$, 
Lucy F. Fortson$^{10}$,
Dale D. Kocevski$^{11}$, 
David C. Koo$^{1,9}$,
Seppo Laine$^{12}$, 
Chris Lintott$^{13,14}$, 
Karen L. Masters$^{15,16}$,
Thomas Melvin$^{15}$,
Robert C. Nichol$^{15,16}$, 
Kevin Schawinski$^{17}$,
Brooke Simmons$^{13}$,
Rebecca Smethurst$^{13}$,
Kyle W. Willett$^{10}$
\vspace{0.1in} }\\
$^{1}$Department of Astronomy and Astrophysics, 1156 High Street, University of California, Santa Cruz, CA 95064\\
$^{2}$Kavli Institute for the Physics and Mathematics of the Universe (WPI), Todai Institutes for Advanced Study, The University of Tokyo\\
$^{3}$Department of Astronomy and Astrophysics, 525 Davey Lab, Penn State University, University Park, PA 16802\\
$^{4}$Aix Marseille Universit\'e, CNRS, LAM (Laboratoire d'Astrophysique de Marseille) UMR 7326, 13388, Marseille, France\\
$^{5}$School of Physics and Astronomy, The University of Nottingham, University Park, Nottingham NG7 2RD, UK\\
$^{6}$Department of Astronomy, University of Michigan, 500 Church St., Ann Arbor, MI 48109, USA\\
$^{7}$The Harriet W. Sheridan Center for Teaching and Learning, Brown University, Box 1912, 96 Waterman Street, Providence, RI 02912, USA\\
$^{8}$Institut de Cincies del Cosmos. Universitat de Barcelona (UB-IEEC), Mart i Franqus 1, E-08028 Barcelona, Spain\\
$^{9}$UCO/Lick Observatory, Department of Astronomy and Astrophysics, University of California, 1156 High Street, Santa Cruz, CA 95064\\
$^{10}$Minnesota Institute for Astrophysics, School of Physics and Astronomy,
University of Minnesota, MN 55455, USA\\
$^{11}$Department of Physics and Astronomy, University of Kentucky, Lexington, KY 40506, USA\\
$^{12}$Spitzer Science Center, MS 314-6, California Institute of Technology, 1200 East California Blvd, Pasadena, CA 91125, USA\\
$^{13}$Oxford Astrophysics, Department of Physics, University of Oxford, Denys Wilkinson Building, Keble Road, Oxford OX1 3RH\\
$^{14}$Astronomy Department, Adler Planetarium and Astronomy Museum, 1300 Lake Shore Drive, Chicago, IL 60605, USA\\
$^{15}$Institute of Cosmology \& Gravitation, University of Portsmouth, Dennis Sciama Building, Portsmouth, PO1 3FX, UK\\
$^{16}$SEPnet, South East Physics Network\\
$^{17}$Institute for Astronomy, Department of Physics, ETH Zurich, Wolfgang-Pauli-Strasse 27, CH-8093 Zurich, Switzerland\\
$^{18}$Hubble Fellow\\
  }
\maketitle
  
\label{firstpage}
  
\begin{abstract}
We present a new study investigating whether active galactic nuclei (AGN) beyond the local universe are preferentially fed via large-scale bars. Our investigation combines data from {\it Chandra} and Galaxy Zoo: {\it Hubble} (GZH) in the AEGIS, COSMOS, and GOODS-S surveys to create samples of face-on, disc galaxies at $0.2 < z < 1.0$. We use a novel method to robustly compare a sample of 120 AGN host galaxies, defined to have $10^{42} ~{\rm erg~s^{-1}} < L_{\rm X} < 10^{44} ~\rm erg~s^{-1}$, with inactive control galaxies matched in stellar mass, rest-frame colour, size, S\'ersic index, and redshift.  Using the GZH bar classifications of each sample, we demonstrate that AGN hosts show no statistically significant enhancement in bar fraction or average bar likelihood compared to closely-matched inactive galaxies. In detail, we find that the AGN bar fraction cannot be enhanced above the control bar fraction by more than a factor of two, at 99.7\% confidence. We similarly find no significant difference in the AGN fraction among barred and non-barred galaxies. Thus we find no compelling evidence that large-scale bars directly fuel AGN at $0.2<z<1.0$. This result, coupled with previous results at $z=0$, implies that moderate-luminosity AGN have not been preferentially fed by {\bf large-scale} bars since $z=1$. Furthermore, given the low bar fractions at $z>1$, our findings suggest that large-scale bars have likely never directly been a dominant fueling mechanism for supermassive black hole growth.
  \end{abstract}
  

%
%
\section{Introduction} \label{sec:introduction}
%
%
\setcounter{footnote}{17}

Most simulations of galaxy evolution require some kind of feedback that correlates with bulge mass (and is often assumed to be active galactic nucleus [AGN] feedback) to reproduce key observations, such as the colour bimodality of galaxies \citep[e.g.,][]{springel05, croton06, cimatti13}. Yet, the mechanism that funnels gas toward the central supermassive black hole that powers the AGN is still unknown (e.g., \citealt{hopkins06}, \citealt{hopkins11}, \citealt{hopkins13}; see \citealt{fabian12}, \citealt{kormendy13}, and \citealt{heckman14} for recent reviews). 

Major mergers are often cited as a key trigger for AGN activity \citep{sanders88, barnes91, mihos96, dimatteo05, hopkins05a, hopkins05b}. Although major mergers seem to drive the most luminous and rapidly accreting AGN \citep{sanders88, koss10, kartaltepe10, treister12, hopkins13, trump13}, low- to moderate-luminosity AGN, which make up the majority of AGN by number, seem to be fueled by processes that do not visibly disturb the discy structure of galaxies \citep{schawinski10, cisternas11, schawinski11, simmons12, kocevski12, schawinski12, simmons13}.   

An obvious process that satisfies this constraint is secular evolution \citep{kormendy79, martinet95, kk04, athanassoula13a, sellwood14}. A major driver of secular evolution in disc galaxies is large-scale bars\footnote{Unless otherwise stated, we use ``bars'' to refer to large-scale
structures in isolated systems, i.e., we do not consider
bars created through interactions. These large-scale bars are 
commonly referred to as primary bars while small-scale (less 
than, or of the order of 1 kpc) bars are commonly referred to 
as secondary bars.}, and they are predicted to affect galaxies in a variety of ways, including the fueling of AGN \citep{simkin80, noguchi88, shlosman89, shlosman90, wada92}. The non-axisymmetric potential of a bar is predicted to funnel interstellar gas into the central kpc \citep{athanassoula92}---which has been confirmed by multiple observational works \citep{regan95, regan99a, sakamoto99, sheth00, sheth02, sheth05, zurita04}---where a possible nested, secondary bar may further funnel gas to the inner $\sim10$ pc. From this distance, cloud-cloud collisions may lead to inflows onto the AGN accretion disc. Collectively, this scenario is known as ``bars within bars'' \citep{shlosman89, shlosman90, hopkins10, hopkins11}. Observations at low redshift, however, find no excess of primary bars in active galaxies (\citealt{ho97, mulchaey97, malkan98, hunt99, regan99b, martini99, erwin02, martini03, lee12b, cisternas13}, but see \citealt{knapen00, laine02, laurikainen04, oh12, alonso13}; Galloway et al. 2014, in preparation). There is also no direct correlation between primary bars and secondary bars, with $\sim30\%$ of all disc galaxies having a secondary bar \citep{mulchaey97, regan99b, martini99, erwin02, laine02}. These results indicate that bars---both primary and secondary bars---may not fuel AGN.

Almost all previous observational work on the bar-AGN connection has been limited to the local universe, where the number density of AGN is low. Thus a compelling link between bars and AGN might still be found at \emph{earlier} epochs, when the number density of AGN is higher  \citep{ueda03, silverman08, aird10}. In this work we focus on galaxies at $0.2<z<1.0$. The upper limit of $z=1$ is based on \cite{melvin14}, who show that bars are detectable out to $z=1$ with the {\it Hubble Space Telescope} Advanced Camera for Surveys ({\it HST}/ACS).

We describe the data in \S2. Our sample selection, and in particular, our selection of control samples of inactive galaxies is detailed in \S3. \S4 presents our main result that there is no statistically significant excess of bars in AGN hosts. We discuss the implications of our results in \S5. Conclusions follow in \S6. Throughout this paper, we assume a flat cosmological model with $H_{0} = 70$ km s$^{-1}$ Mpc$^{-1}$, $\Omega_{m} = 0.30$, and  $\Omega_{\Lambda} =0.70$, and all magnitudes are given in the AB magnitude system.

%
%
\section{Data}\label{sec:data}
%
%

In this section, we briefly describe the three surveys and their respective data products that are used in this paper. We also describe Galaxy Zoo: {\it Hubble}, which uses high resolution {\it HST}/ACS imaging to accurately visually classify galaxies. Thus this paper will only focus on the area of these surveys that have {\it HST}/ACS imaging. A summary of the three surveys is presented in Table~\ref{tab:survey_summary}.

\begin{table}
\begin{center}
\begin{minipage}[c]{8cm}
        \caption{Survey Summary}
         \label{tab:survey_summary}
        \begin{tabular}{cccc}
             & AEGIS  & COSMOS & GOODS-S \\
        \hline
        \hline
Area (deg$^2$) & 0.197 & 1.8 & 0.07 \\ 
{\it HST}/ACS Exp Time (s)\footnote{For AEGIS and GOODS-S, this is the average exposure time of the two observed {\it HST}/ACS bands used in this work (see \S\ref{sub:aegis_data} and \S\ref{sub:goodss_data})} & 2180 & 2028 & 2223 \\
Pixel Scale (\arcsec pixel$^{-1}$) & 0.03 & 0.05 & 0.03  \\
PSF FWHM (\arcsec) & 0.120 & 0.090 & 0.125 \\
        \hline
        \end{tabular}
\end{minipage}
\end{center}
\end{table}

\subsection{AEGIS} \label{sub:aegis_data}

The All-wavelength Extended Groth strip International Survey (AEGIS; \citealt{davis07}) is an international collaboration that produced one of the most comprehensive multi-wavelength data sets currently available. This data set includes {\it HST}/ACS imaging, which is centered on the EGS region and is composed of 63 pointings in both the F606W ($V$) and the F814W ($I$) filters. The final images have a pixel scale of $0\farcs03\rm ~pixel^{-1}$ and a point-spread function (PSF) of $0\farcs12$ FWHM. The {\it HST}/ACS images cover a total area of $\sim710~\rm arcmin^2$. 

The multi-wavelength coverage of AEGIS also includes $Chandra$ ACIS-I \citep{garmire03} $X$-ray observations that have a nominal exposure of 800 ks \citep{nandra05, georgakakis06, laird09}.

The spectroscopic redshifts ($z$) of AEGIS are from the DEEP2 and DEEP3 redshift surveys \citep{davis03, newman13, cooper11, cooper12}, which used the DEIMOS spectrograph \citep{faber03} on the Keck II telescope. Spectroscopic redshifts with quality code of 3 or 4 are considered secure; we only consider these spectroscopic redshifts throughout this paper. 

Measurements of AEGIS galaxy properties are from \cite{cheung12}, who compiled several galaxy measurements from several different sources. Stellar masses, $M_*$ are from \cite{huang13}, and are estimated by fitting the multi-wavelength AEGIS photometry to a grid of synthetic SEDs from \cite{Bruzual03}, assuming a \cite{salpeter55} IMF and solar metallicity. These synthetic  SEDs span a range of ages, dust content, and exponentially declining star formation histories (SFHs). 

Rest-frame $U-B$ colours are obtained through the $k$-correct v4.2 code \citep{blanton07} with CFHT $BRI$ photometry and spectroscopic redshifts as inputs. 

Structural parameters such as the global S\'ersic index ($n$), effective radius ($r_{\rm e}$), and axis ratio ($b/a$) are measured with GIM2D through a single S\'ersic fit on the {\it HST}/ACS $V$ and $I$ images \citep{simard02}.

\subsection{COSMOS} \label{sub:cosmos_data}

The Cosmological Evolution Survey (COSMOS; \citealt{scoville07, koekemoer07}) is the largest contiguous {\it HST}/ACS imaging survey to date, covering $\sim1.8~\rm deg^2$ in the F814W ($I$) band and consists of 590 pointings. The final images have a pixel scale of $0\farcs05\rm ~pixel^{-1}$ and a point-spread function (PSF) of $0\farcs09$ FWHM. 

In addition to the {\it HST}/ACS coverage, COSMOS also includes $Chandra$ ACIS-I observations that cover the central part of the COSMOS field with four pointings, each totaling to a nominal exposure of 200 ks \citep{elvis09}. We use the $Chandra$ COSMOS catalog as described in \cite{civano12}.

The spectroscopic $z$'s of COSMOS are mainly from zCOSMOS \citep{lilly09}. Supplemental spectroscopic $z$'s are from the $Chandra$ COSMOS survey \citep[e.g., from][]{trump09}. We only consider spectroscopic $z$'s that are deemed secure by these surveys, e.g., for zCOSMOS, we only consider $z$'s with confidence class of 3.x, 4.x, 1.5, 2.4, 2.5, 9.3, 9.5, 13.x, 14.x, 23.x, and 24.x. 

Measurements of COSMOS galaxy properties are from a variety of sources. Stellar masses and rest-frame $U-V$ colours are from the UltraVISTA survey \citep{muzzin13}. The $M_*$'s are estimated using FAST \citep{kriek09} to fit the galaxy SEDs to \cite{Bruzual03} models, assuming solar metallicity, a \cite{chabrier03} IMF, a \cite{calzetti00} dust extinction law, and exponentially-declining SFHs. 

The rest-frame $U-V$ colours are estimated by using EAZY \citep{brammer08} to determine the colours by integrating the best-fit SED through the redshifted filter curves over the appropriate wavelength range. 

The structural parameters of COSMOS galaxies, i.e., $n$, $r_{\rm e}$, and $b/a$, are provided by the ACS-GC catalog \citep{griffith12}; they used GALFIT \citep{peng02} to fit a single S\'ersic profile on the {\it HST}/ACS $I$ images.

\subsection{GOODS-S} \label{sub:goodss_data}

The Great Observatories Origins Deep Survey (GOODS; \citealt{dickinson03, giavalisco04, rix04}) is a deep multiwavelength survey that includes the deepest {\it HST} images to date. The GOODS survey targeted two separate fields, the {\it Hubble} Deep-Field North (HDF-N; now referred to as GOODS-N) and the $Chandra$ Deep-Field South (CDF-S; now referred to as GOODS-S). We will only use the GOODS-S for this paper. 

The {\it HST}/ACS imaging of GOODS-S was carried out in several bands, of which we are only interested in two -- F606W ($V$) and F850LP ($z$). The imaging comprises of 15 pointings, with a final pixel scale of the images of $0\farcs03\rm ~pixel^{-1}$ and a point-spread function (PSF) of $0\farcs125$ FWHM. The {\it HST}/ACS imaging area of GOODS-S covers a total area of $\sim160~\rm arcmin^2$. 

In addition to containing the deepest {\it Hubble} images to date, GOODS-S also contains the deepest $Chandra$ observations to date. The 4 Ms CDF-S Survey \citep{luo08, xue11} made 54 $Chandra$ ACIS-I observations over a period of three $Chandra$ observing cycles in 2000, 2007, and 2010. We use the catalog presented by \cite{xue11}.

The spectroscopic $z$'s of GOODS-S come from a variety of sources, many of which are listed in Table 2 of \cite{griffith12}. We only consider redshifts of the highest quality ($\ge3$). 

Stellar masses and rest-frame $U-B$ colours of GOODS-S galaxies are from the CANDELS survey \citep{koekemoer11,grogin11,barro11a, guo13, williams14}. The $M_*$'s are estimated with the FAST code by fitting galaxy SEDs based on optical to infrared photometry to models of \cite{Bruzual03}, assuming a \cite{chabrier03} IMF. They also assumed a \cite{calzetti00} extinction law, solar metallicity, and exponentially-declining SFHs. 

Rest-frame $U-B$ colours are estimated with the EAZY code by fitting galaxy SEDs to the templates from \cite{muzzin13}. 

The structural parameters of GOODS-S galaxies, i.e., $n$, $r_{\rm e}$, and $b/a$, are provided by the ACS-GC catalog \citep{griffith12}; they used GALFIT \citep{peng02} to fit a single S\'ersic profile on the {\it HST}/ACS $V$ and $z$ images.

\subsection{Calculating $L_{\rm X}$}

To identify AGN, we use full-band (0.5-10 keV for AEGIS and COSMOS, 0.5-8 keV for GOODS-S)  {\it X}-ray fluxes from {\it Chandra} observations described by \cite{laird09}, \cite{civano12}, and \cite{xue11} for AEGIS, COSMOS, and GOODS-S, respectively. We calculate {\it X}-ray luminosities using the equation $L_{\rm X} = 4\pi d^2_{L}f_{x}(1+z)^{\Gamma -2}$, where $d_{L}$ is the luminosity distance, $z$ is the redshift, $f_{x}$ is the flux, and $\Gamma$ is the power-law photon index. We set $\Gamma=1.8$, which is a typical power-law photon index for intrinsic AGN spectra. In \S\ref{sub:agn_selection}, we select AGN using these {\it X}-ray luminosities.

\begin{figure*}
\centering
\includegraphics[scale=1.1]{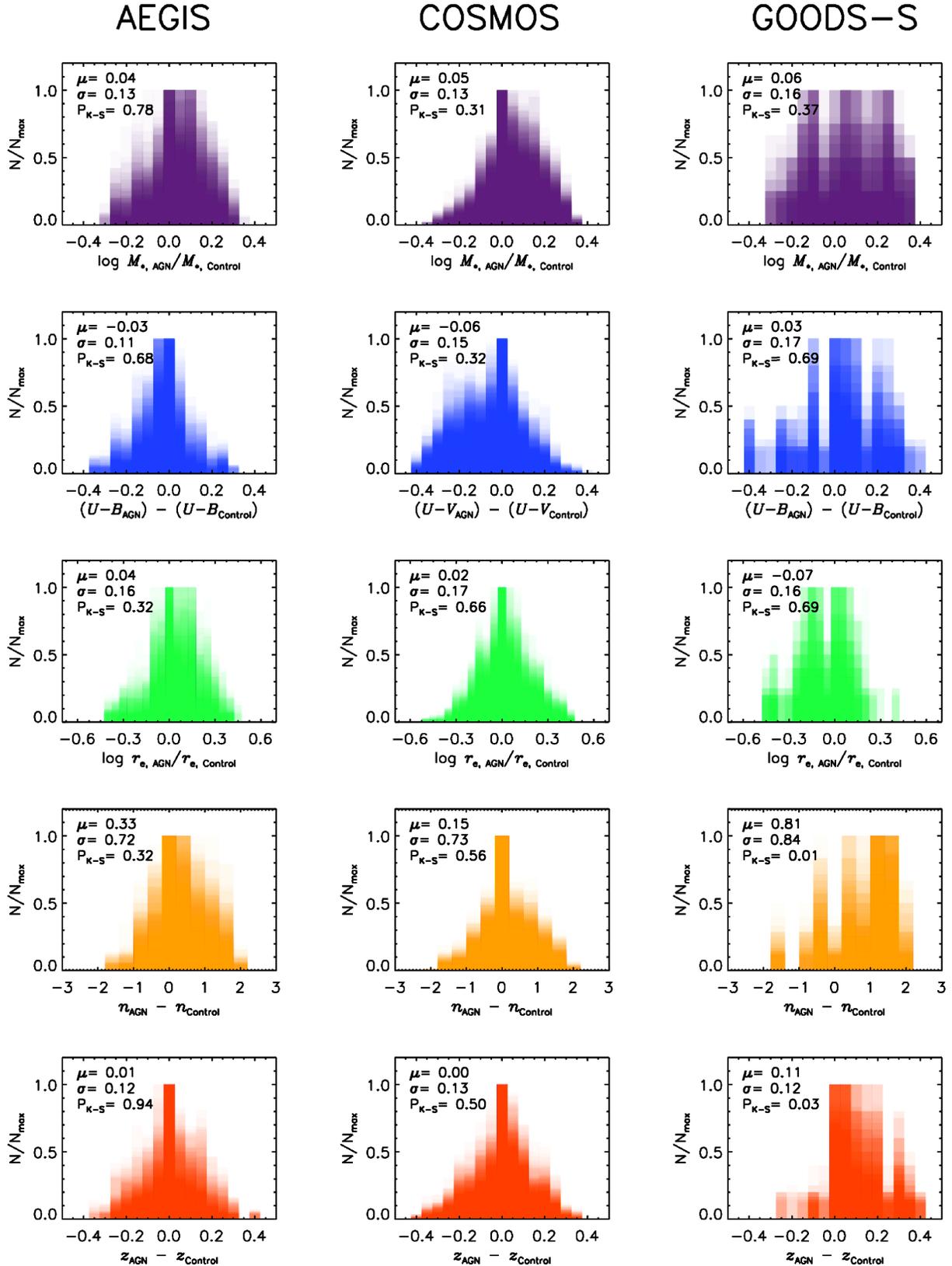}
\caption{
Normalized histograms of differences in matching parameters of the AGN and control samples. From top to bottom, the parameters are: stellar mass ($M_*$), rest-frame colour ($U-B ~{\rm or}~ U-V$), effective radius ($r_{\rm e}$), global S\'ersic index ($n$), and spectroscopic redshift ($z$). Each panel overlays 100 histograms, one for each AGN-control sample realization; the mean ($\mu$), standard deviation ($\sigma$), and the K-S null probability (P$_{\rm K-S}$) is displayed in the upper left of each panel. The shading reveals the amount of overlap. Most of the histograms peak around zero, implying that the AGN and control galaxies are generally well-matched. The histograms of GOODS-S are slightly broader and more skewed than those of AEGIS and COSMOS, which is due to the relatively small sample size of GOODS-S. 
 \label{fig:agn_controls}}
\end{figure*}

\begin{figure*}
\centering
\includegraphics[width=\textwidth]{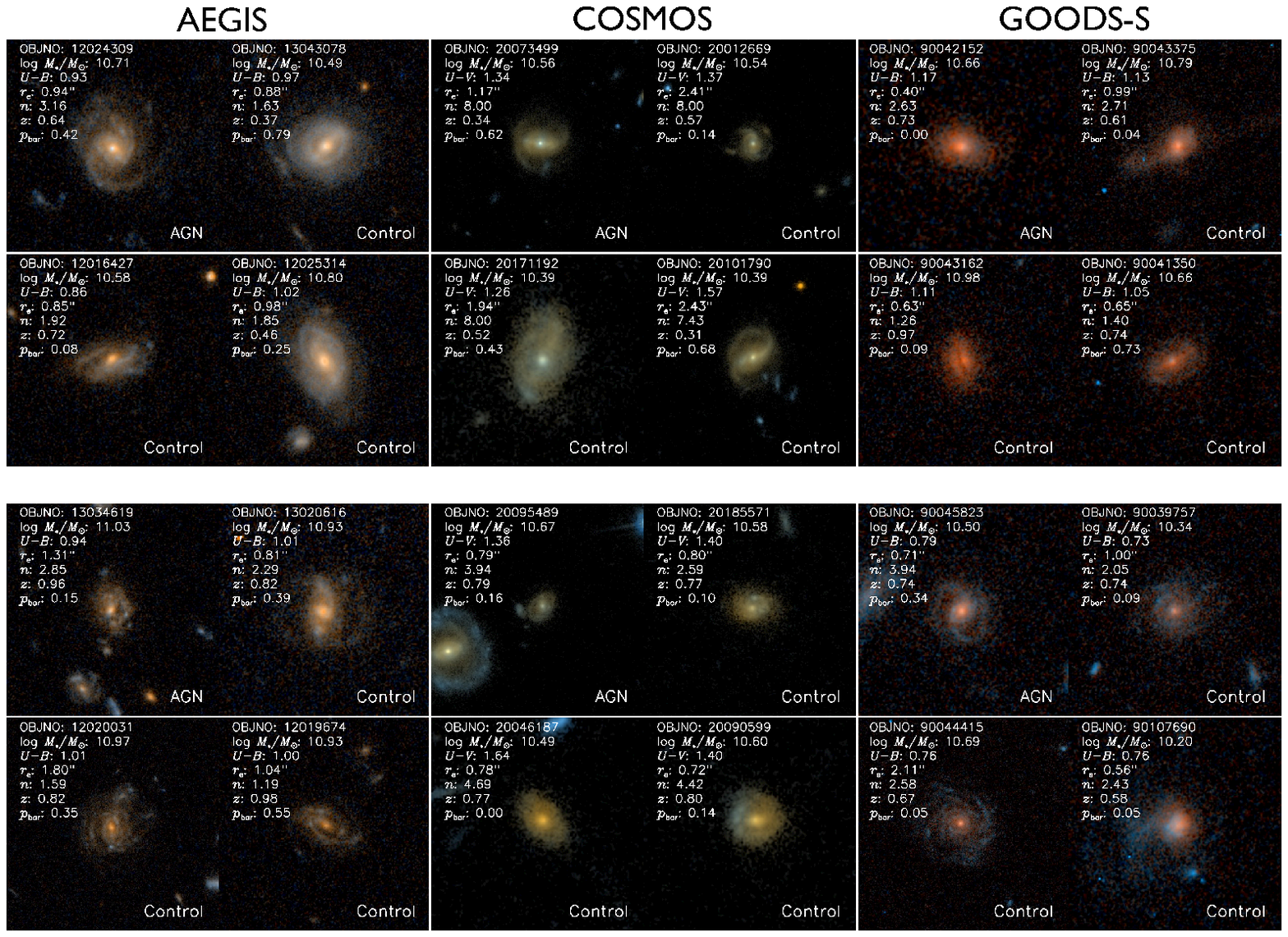}
\caption{
{\it HST}/ACS images of two matched sets of AGN-control galaxies for AEGIS (left), COSMOS (middle), and GOODS-S (right). Each AGN (upper-left image of each block) has three matched, non-AGN control galaxies. Galaxies with $p_{\rm bar}>0.5$ are considered barred. Galaxies within each AGN-control set is similar in appearance, demonstrating the quality of our matching technique. Images are from Griffith et al. 2012.
\label{fig:agnbar_gallery}}
\end{figure*} 

\subsection{Galaxy Zoo: {\it Hubble}} \label{sub:galaxyzoohubble_data}

Our work relies on bar identifications from the Galaxy Zoo: {\it Hubble} citizen science project (GZH; \citealt{melvin14}). Volunteers were asked to visually classify the morphologies of galaxies at $z\sim1$ based on {\it HST}/ACS optical imaging from the surveys listed above. Like the previous Galaxy Zoo project, Galaxy Zoo 2 \citep{willett13}, GZH used a decision tree with multiple branches and nested, dependent questions.\footnote{The complete decision tree is available at \href{http://data.galaxyzoo.org}{http://data.galaxyzoo.org}.} These questions include, {\it ``Is there a sign of a bar feature through the centre of the galaxy?''}, which can only be reached if a volunteer identifies some type of a feature (e.g., clumps, spiral arms, rings, bars) or a disc within a galaxy. Thus a galaxy must have a feature or a disc in order to be classified as barred. 

Each galaxy is classified by at least 33 volunteers, with the median number of volunteers per galaxy being 47. These classifications produce vote percentages that we refer to throughout as ``likelihoods''; e.g., if 25 out of 50 volunteers classified a galaxy as having a bar, then the bar likelihood is $p_{\rm bar}=0.5$, moduli small corrections to downweight consistently unreliable classifiers (following the procedure explained in \citealt{willett13}).

GZH only classifies galaxies brighter than the following magnitudes: for AEGIS and COSMOS,  {\it HST}/ACS F814W $< 23.5$ (AB) magnitude, and for GOODS-S, {\it HST}/ACS F850LP$ < 23.5$ (AB) magnitude.

\subsubsection{Selecting Barred Galaxies} \label{sub:bar_selection}

We consider galaxies to be barred if they have bar likelihoods greater than 0.5 ($p_{\rm bar}\ge0.5$) and no obvious dust lanes\footnote{The exclusion of this criterion does not significantly affect the measured bar fractions, nor does it change our conclusions.} ($p_{\rm dust~lane}<0.5$). The bar threshold of $p_{\rm bar}=0.5$ is based on previous Galaxy Zoo works that have shown it to be a reliable indicator of strong bar features \citep{masters11, masters12, willett13, melvin14}. 

To calculate the bar fraction, $f_{\rm bar}$, one can divide the total number of barred galaxies by the total number of disc galaxies in the sample. Varying the bar likelihood threshold between $0.3 \leq  p_{\rm bar} \leq 0.6$ does not change our qualitative conclusions. 

In our analysis, we also use the average bar likelihood, $\overline{p}_{\rm bar}$, defined as the average of all the bar likelihoods of a sample of galaxies. This parameter is another measure of bar presence and has been used by other studies \citep[e.g.,][]{skibba12, casteels13, cheung13}.

\section{Sample Selection} \label{sec:selection}

Since our study seeks to determine if AGN activity is linked with bars, we first construct a parent sample of face-on, disc-dominated objects whose bars can be robustly identified (\S\ref{sub:disc_selection}). We then select AGN-hosting galaxies from this parent sample based on their {\it X}-ray luminosity (\S\ref{sub:agn_selection}). Finally, we construct samples of inactive control galaxies that are matched to the AGN galaxies (\S\ref{sub:control_selection}). The number counts of these samples are listed in Table \ref{tab:sample_counts}.

\subsection{Face-on Disc Selection} \label{sub:disc_selection}

The face-on disc samples for each field are defined by the following criteria:

\begin{enumerate}
\item $0.2<z<1.0$ -- To obtain the most accurate {\it X}-ray luminosities and to identify broad-line AGN (which may contaminate their host galaxy measurements), we choose only galaxies with secure spectroscopic redshifts. Although the ability to identify a bar is not uniform over this redshift range, our robust matching of AGN and inactive control galaxies ensures that the two samples have the same distributions of completeness for bar detection: see \S\ref{sub:redshift_effects}.
\item $b/a > 0.5$ -- Since bars in highly inclined galaxies are difficult to identify, we exclude edge-on galaxies with global axis ratios less than or equal to 0.5. 
\item $r_{e} > 8 ~\rm pixels$ -- Selecting galaxies with $r_{\rm e}$ larger than 8 pixels, which corresponds to about twice the FWHM of the {\it HST}/ACS PSF, ensures that any bars with semimajor axes $\gs 3$ kpc will be identified. Since the typical bar lengths in the local universe are $2-7$ kpc \citep{erwin05, gadotti11, hoyle11}, we should be able to detect most large-scale barred galaxies, assuming bars at $z>0$ are similar to bars at $z\sim0$.  
\item $N_{\rm Bar ~question}/N_{\rm Total} \ge 0.15$ -- In order to answer the bar question, the GZH decision tree requires a volunteer to classify a galaxy as displaying some kind of feature or disc. Thus demanding that at least $15\%$ of a galaxy's classifiers answer the bar question results in an effective selection for disc galaxies. Although the number of AGN is sensitive to the exact $N_{\rm Bar ~question}/N_{\rm Total}$ threshold, our qualitative conclusions are not, e.g., requiring $N_{\rm Bar ~question}/N_{\rm Total} \ge 0.75$ does not change our ultimate conclusion. Moreover, the majority of our $N_{\rm Bar ~question}/N_{\rm Total} \ge 0.15$ sample has more than ten bar classifications, which is more than most visual bar classifications \citep[e.g.][]{nair10b, lee12a}.
\item $p_{\rm merge} < 0.65$ -- In order to separate out the effects of mergers from our analysis, we choose non-interacting galaxies by requiring a merging likelihood ($p_{\rm merge}$) less than 0.65. This criterion mirrors that of \cite{melvin14} and eliminates a small fraction of our sample. Discarding this criterion does not affect our conclusion.
\end{enumerate}

\begin{figure*}
\centering
\includegraphics[scale=.85]{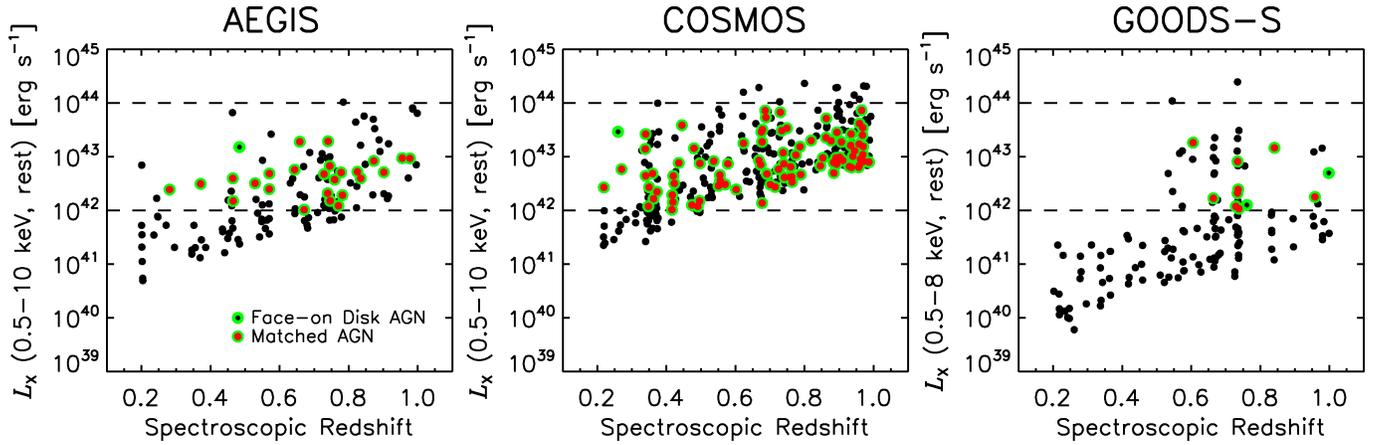}
\caption{{\it X}-ray luminosity vs. spectroscopic redshift for all sources detected in the AEGIS (left), COSMOS (middle), and GOODS-S (right) surveys in our chosen redshift range. The dashed horizontal lines represent the lower and upper limits of our AGN selection. Our AGN samples, i.e., the red points encircled in green, span most of $L_{\rm X}-z$ space.
\label{fig:lx_redshift_all}}
\end{figure*} 

\subsection{AGN Selection} \label{sub:agn_selection}

Out of these face-on disc samples (one from each survey) we select AGN hosts with {\it X}-ray luminosities $10^{42} ~{\rm erg~s^{-1}} < L_{\rm X} < 10^{44} ~\rm erg~s^{-1}$. The lower limit removes starburst galaxies with weak {\it X}-ray emission \citep{bauer02}, and the upper limit excludes quasars that may have optical point sources which would contaminate the measurements of their host galaxies \citep{silverman08b}.

We also discard luminous unobscured AGN which might also contaminate the visible appearance of their host galaxy measurements. In AEGIS and COSMOS, we use broad emission lines to identify such AGN since they dominate the optical spectra of their host galaxies. There are zero broad-line AGN in AEGIS and five broad-line AGN in COSMOS, which we reject from our sample.  GOODS-S lacks a public catalog of broad-line AGN, and so we instead use low {\it X}-ray hardness (HR $<-0.3$; \citealt{mainieri07}) as a proxy for unobscured AGN: this results in the rejection of one AGN.  The rejection or inclusion of these six potential-contaminant AGN does not affect our qualitative conclusions.

Table \ref{tab:sample_counts} lists the final AGN count for each survey. 

\subsection{Control Selection}  \label{sub:control_selection}

For each AGN host galaxy, we select three unique, non-AGN control galaxies from the \emph{same} survey (AEGIS, COSMOS, or GOODS-S) that are matched in $M_*$, $U-B$\footnote{$U-V$ for COSMOS}, $n$, $r_{\rm e}$, and $z$. These parameters have been shown to correlate with both AGN presence and bar presence \citep[e.g.,][]{kauffmann03, nandra07, schawinski09, masters11, cheung13} and thus must be controlled for in order to uncover any underlying bar-AGN connection.

For a given AGN host galaxy, we first select a pool of control galaxies satisfying the following conditions: 

\begin{itemize}
\item $\lvert \log ~M_{*, \rm AGN}/M_* \rvert < 0.35$
\item $\lvert (U - B)_{\rm AGN} - (U - B) \rvert < 0.4$
\item $\lvert \log ~r_{\rm e, AGN}/r \rvert < 0.48$
\item $\lvert n_{\rm AGN} - n \rvert < 2.0$
\item $\lvert z_{\rm AGN} - z \rvert < 0.4$
\end{itemize}

These limits are tuned in order to find enough control galaxies for each AGN. Our conclusions are not sensitive to the exact limits, e.g., reducing these limits by $50\%$ does not change the conclusions.

With this initial pool of control galaxies, we perform a 5-stage matching process that iteratively reduces the pool until it reaches a final set of three unique and matched control galaxies. The first stage cuts the initial pool of control galaxies to the 15 closest matched galaxies in one of the matching parameters, i.e., $M_{*, \rm AGN}$, $U-B_{\rm AGN}$\footnotemark[21], $n_{\rm AGN}$, $r_{\rm e, AGN}$, or $z_{\rm AGN}$. Each successive stage matches the remaining control galaxies to one of the unused matching parameters and eliminates the three worst matched galaxies. Thus the second stage reduces the pool to 12, the third stage reduces the pool to 9, the fourth reduces the pool to 6, and finally, the fifth stage reduces the pool to 3. Ultimately, for each survey we have a control sample that contains no duplicates and is three times larger than the AGN sample (see Table \ref{tab:sample_counts}). 

This 5-stage matching process was performed for each AGN host in our sample.  However, there were four AGN---one from AEGIS, one from COSMOS, and two from GOODS-S---that were discarded due to a lack of control galaxies for the AGN host.  These host galaxies have abnormally high $n$ for their rest-frame colours and/or $M_*$ compared to the pool of control galaxies.  

The AGN and control samples that this matching technique produces are affected by the order in which we match the AGN hosts to the control galaxies and the order in which the matching parameters are used. In order to adequately sample the parameter space, we repeat this 5-stage matching technique 100 times for each survey, with each iteration randomly shuffling the order of the AGN hosts, the order of the control galaxies, and the order of the matching parameters. Ultimately, we generate 100 AGN samples and 100 control samples for each survey. For brevity and clarity, we define an ``AGN-control sample'' to be all AGN hosts and their corresponding control galaxies for a given survey and for a given matching iteration. We use the median counts and the resulting $f_{\rm bar}$ and $\overline{p}_{\rm bar}$ of the 100 AGN-control samples in presenting our results (see Table \ref{tab:sample_counts}). 

To demonstrate the quality of our matching procedure, Fig.~\ref{fig:agn_controls} presents stacked histograms of the distribution of parameter differences between the AGN hosts and their matched control galaxies. Each panel stacks 100 translucent histograms, with each histogram representing one AGN-control sample. To elaborate, each histogram represents the difference in a parameter between each AGN and its three control galaxies for a given realization. The highly shaded regions represent the most populated parameter space, which are generally centered around 0 with small spreads, as supported by the mean and standard deviation at the upper left of each panel, indicating that our matching technique works well. Moreover, we also calculated the two-sample Kolmogorov-Smirnov (K-S) null probability for each pair of AGN-control parameter distributions, where small values indicate that the two distributions in question are probably not from the same underlying distribution. We display the median K-S null probability, P$_{\rm K-S}$, of all 100 pairs of AGN-control distributions in each panel, most of which show high values, indicating that the AGN and control samples are consistent.

However, it is clear that the histograms of GOODS-S are slightly broader and more skewed than those of AEGIS and COSMOS, especially in S\'ersic index and redshift, as supported by P$_{\rm K-S}$. The relatively small sample size of GOODS-S (see Table \ref{tab:sample_counts}) makes it difficult to identify well-matched control galaxies for the GOODS-S AGN hosts. However, \S\ref{sec:results} shows that the results from the GOODS-S sample are consistent with those of AEGIS and COSMOS, indicating that the skewness of the GOODS-S AGN-control samples does not bias our analysis.

\begin{figure*}
\centering
\includegraphics[scale=.78]{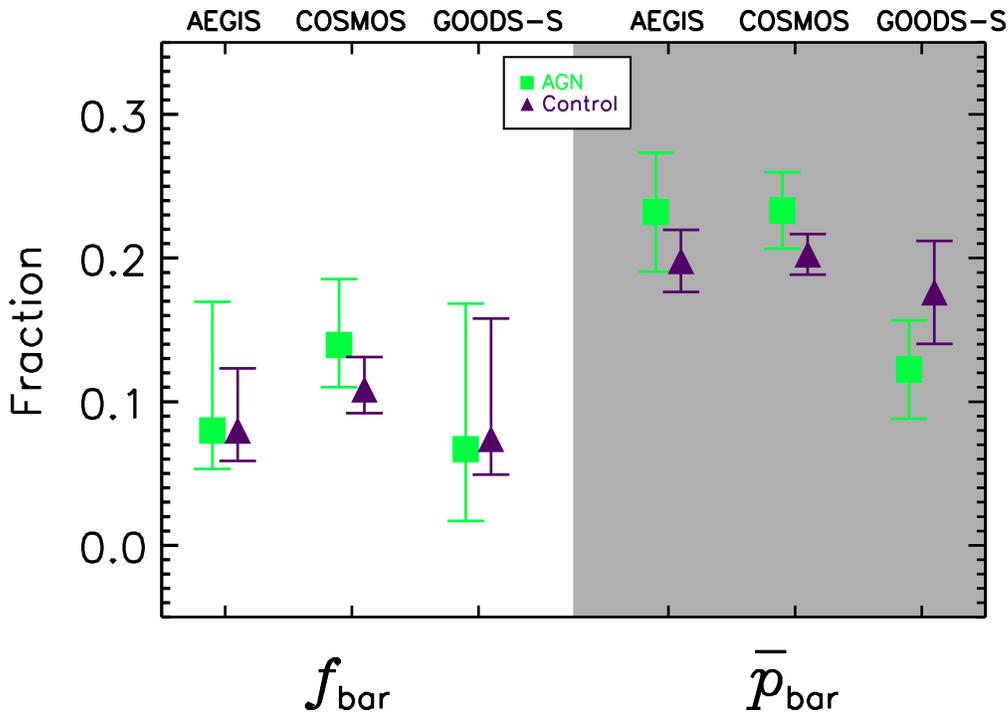}
\caption{{\it Left}: The bar fraction, $f_{\rm bar}$, of the AGN (green squares) and non-AGN control samples (purple triangles) for the AEGIS, COSMOS, and GOODS-S surveys. {\it Right}: The average bar likelihood, $\overline{p}_{\rm bar}$, of the AGN and non-AGN control samples for the three surveys. The error bars on $f_{\rm bar}$ and $\overline{p}_{\rm bar}$ are the $68.3\%$ binomial confidence limits and standard error, respectively. There is no statistically significant difference in $f_{\rm bar}$ or $\overline{p}_{\rm bar}$ between the AGN and non-AGN control samples across all three surveys, indicating that there is no large excess of bars in AGN hosts.
\label{fig:bar_agn_frac}}
\end{figure*} 

To further illustrate the quality of our matching technique, we show images of two matched sets of AGN-control galaxies from each survey in Fig.~\ref{fig:agnbar_gallery}. Each set of AGN-control galaxies is reassuringly similar in appearance, confirming that our matching technique is reasonable.

The {\it X}-ray luminosity-redshift distribution of all {\it X}-ray sources in our chosen redshift range, i.e., from the first row of Table \ref{tab:sample_counts} labelled ``$0.2<z<1.0$'', is shown in Fig.~\ref{fig:lx_redshift_all}. The dashed horizontal lines define our AGN selection. The black points encircled in green represent the AGN that satisfy our face-on disc criteria, and hence are eligible to undergo our matching process. The red points encircled in green represent the face-on disc AGN that have enough control galaxies, and thus are in our AGN samples. Fig.~\ref{fig:lx_redshift_all} shows that our AGN samples span most of $L_{\rm X}$-$z$ space. It also shows that the number of moderate-luminosity AGN increases with $z$, a well-known behaviour that has been shown by previous works \citep[e.g.,][]{ueda03, silverman08, aird10}.

\subsection{Redshift Effects on Bar Detection} \label{sub:redshift_effects}

Although redshift effects (e.g., cosmological surface brightness dimming, angular size change, band-shifting) hinder bar detection, our experiment is a \emph{relative} comparison between two matched samples that controls for redshift and several of the most correlated parameters of bar presence (e.g., stellar mass, colour, and S\'ersic index; \citealt{nair10b, masters11, lee12a, cheung13}), so our study naturally takes this bias into account. Assuming that there are no AGN-dependent selection effects, our experiment should be robust against any known bar detection biases. 

As an additional check, we tested for differential redshift effects by splitting our sample into two redshift intervals, $0.20<z<0.84$ and $0.84<z<1.00$, and repeating our analysis. We chose these $z$ intervals because \cite{sheth08} argued that the ability to detect bars at $z>0.84$ is hampered by the overlap of the {\it HST}/ACS $I$ band with the rest-frame near-UV, where clumpy star formation can hide smooth bar structures. However, as pointed out by \cite{melvin14}, the majority of the light gathered in the {\it HST}/ACS $I$ band at $0.84 < z < 1.00$ is from the rest-frame optical, and it is only beyond $z\sim1$ that this filter becomes dominated by rest-frame UV where bar detection may be hindered. Therefore we do not expect a reduction in our ability to detect bars above $z=0.84$. Repeating our analysis for these two $z$ intervals corroborates our expectations: there is no statistically significant ($<3\sigma$) difference in the AGN sample's bar fraction and the control sample's bar fraction at either $0.20<z<0.84$ or at $0.84<z<1.00$. Therefore we find no redshift dependence on our results, indicating that our results are not affected by redshift biases.  

\section{Results} \label{sec:results}

\subsection{Do AGN Hosts Contain An Excess Of Large-scale Bars?}

\begin{table}
\begin{minipage}[c]{8cm}
  \begin{threeparttable}
    \caption{Sample Statistics}
     \label{tab:sample_counts}
     \begin{tabular}{cccc}
        \toprule
             & AEGIS  & COSMOS & GOODS-S \\
        \midrule
$0.2<z<1.0$ & 3,958 & 6,673 & 1,023\\
Face-on Disc & 1,227 & 2,244 & 260 \\ \noalign{\smallskip} \hline \noalign{\smallskip}
AGN & 25 & 86 & 9 \\  
Control &  75  & 258 & 27 \\
Barred AGN & 2 & 12 & 0 \\ 
Barred Control & 6 & 28 & 2 \\  \noalign{\smallskip} \hline \noalign{\smallskip} \vspace{-2mm}
$f_{\rm bar,~AGN}$  &  $0.08\substack{+0.09 \\ -0.03} $  & $0.14\substack{+0.05 \\ -0.03} $ & $0.07\substack{+0.10 \\ -0.05} $ \footnote{According to \cite{cameron11}, when $f_{\rm bar}=0$, one can adopt the median of the beta distribution likelihood function as one's best guess for the true $f_{\rm bar}$.}\\ \\ \vspace{-2mm}
$f_{\rm bar,~Control}$  &  $0.08\substack{+0.04 \\ -0.02}$ & $0.11\substack{+0.02 \\ -0.02} $ & $0.07\substack{+0.08 \\ -0.03} $\\ \\ \vspace{-2mm}
$\overline{p}_{\rm bar,~AGN}$  & $0.23\pm{0.04}$ &  $0.23\pm{0.03}$ & $0.12 \pm{0.03}$ \\ \\
$\overline{p}_{\rm bar,~Control}$ & $0.20\pm{0.02}$ &  $0.20\pm{0.01}$  & $0.18\pm{0.04}$ \\ 
        \bottomrule
     \end{tabular}
    \begin{tablenotes}
      \small
      \item The first row represents the total \# of galaxies at $0.2<z<1.0$ with secure spectroscopic $z$'s and {\it HST}/ACS imaging in our sample, and the second row represents the \# of face-on disc galaxies that are in the previous row. The rest of the table shows the median counts from the 100 AGN-control samples (see \S3.3), and the resulting bar fraction, $f_{\rm bar}$, and average bar likelihood, $\overline{p}_{\rm bar}$. 
    \end{tablenotes}
  \end{threeparttable}
  \end{minipage}
\end{table}

The main result of this paper is shown in Fig.~\ref{fig:bar_agn_frac}, which plots the bar fraction, $f_{\rm bar}$, and the average bar likelihood, $\overline{p}_{\rm bar}$, of the AGN and non-AGN control samples for the AEGIS, COSMOS, and GOODS-S surveys. The uncertainties shown for $f_{\rm bar}$ are $68.3\%$ binomial confidence limits, calculated using quantiles of the beta distribution given the bar counts, total sample counts, and desired confidence level\footnote{For small samples, one can refer to the reference Tables in \cite{cameron11}.} \citep{cameron11}. The uncertainties shown for $\overline{p}_{\rm bar}$ are calculated as $\sigma/\sqrt{N}$, where $\sigma$ is the standard deviation of $p_{\rm bar}$, and $N$ is the total number of galaxies. 

We find no statistically significant enhancement in $f_{\rm bar}$ or $\overline{p}_{\rm bar}$ in AGN hosts compared to the non-AGN control galaxies. The probabilities that each survey's $f_{\rm bar, AGN}$ and $f_{\rm bar, Control}$ are different, given the binomial errors, are insignificant ($\ls$ 1$\sigma$). Conducting a two-sample K-S test on the $p_{\rm bar}$ distributions of each survey's AGN and control samples reveals that the AGN and control samples are consistent with being drawn from the same parent sample to the $99.9\%$ level. 

With our results, we can quantify the level of bar excess in AGN hosts that we can eliminate by combining all three surveys together. Using the combined counts of AGN, controls, barred-AGN, and barred-controls, we find that the bar fraction of the combined AGN sample cannot be greater than twice the bar fraction of the combined control sample at 99.7\% confidence. Therefore, we conclude that \emph{there is no large excess of bars in AGN hosts}. 

\subsection{Are Large-scale Bars Efficient Fuelers Of AGN?}

\begin{figure}
\centering
\includegraphics[scale=.49]{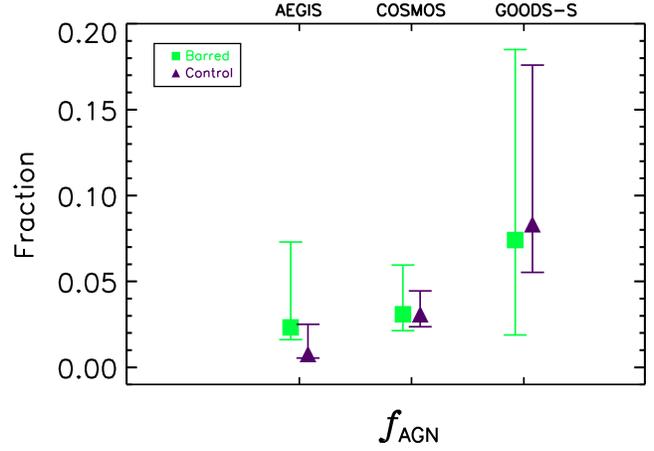}
\caption{The AGN fraction, $f_{\rm AGN}$, of the barred (green squares) and non-barred control samples (purple triangles) for the AEGIS, COSMOS, and GOODS-S surveys. The error bars on $f_{\rm AGN}$ are the $68.3\%$ binomial confidence limits. There is no statistically significant difference in $f_{\rm AGN}$ between the barred and non-barred control samples across all three surveys, indicating that there is no statistically significant excess of AGN in barred galaxies.}
\label{fig:agn_fraction}
\end{figure}

A slightly different, but related question is, ``Are bars efficient fuelers of AGN?'' We answer this question by studying the AGN fraction of barred and non-barred galaxies, as presented in Fig.~\ref{fig:agn_fraction}. From our face-on disc sample (see \S\ref{sub:disc_selection}), we select barred galaxies with the criteria described in \S\ref{sub:bar_selection}. We then create a control sample of non-barred galaxies by demanding that $p_{\rm bar}<0.05$ and by using our 5-stage matching technique that we described in \S\ref{sub:control_selection}. That is, for each barred galaxy, we find three non-barred galaxies matched in stellar mass, rest-frame colour, size, S\'ersic index, and redshift. Using the AGN criteria defined in \S\ref{sub:agn_selection}, Fig.~\ref{fig:agn_fraction} shows that there is no statistically significant excess of AGN among barred galaxies. 

\section{Discussion} \label{sec:discussion}

At $z\sim0$, several works have previously found no link between bars and AGN (\citealt{ho97, mulchaey97, malkan98, hunt99, regan99b, martini99, erwin02, martini03, lee12b, cisternas13}, but see \citealt{laine02, knapen00, laurikainen04, oh12,  alonso13}; Galloway et al. 2014, in preparation), and our results suggest that this absence of direct bar-driven AGN activity persists out to $z=1$. Our chosen redshift range corresponds to an epoch where approximately half of the local supermassive black hole mass density was formed \citep{aird10}, indicating that bars are not directly responsible for the buildup of at least half of the local supermassive black hole mass density. Moreover, the paucity of bars at $z>1$ \citep{kraljic12, simmons14} indicates that bars were probably not closely associated with AGN at $z>1$ either. Therefore, large-scale bars are likely not the primary fueling mechanism for supermassive black hole growth over cosmic time.  

Recently, \cite{cisternas14} also searched for a bar-AGN connection using a slightly smaller sample of 95 AGN in COSMOS, with both photometric and spectroscopic redshifts at $0.15<z<0.84$.  Their results are broadly consistent with ours, with neither a significant bar excess among AGN nor an AGN excess among barred galaxies at $z>0.4$. However, \cite{cisternas14} do suggest a marginal excess of bars among AGN (compared to non-AGN) at $z \sim 0.3$, which we do not detect in our results. This difference is likely due to slight differences in bar identification\footnote{The discrepancy in GZH bar fractions and those used by \cite{cisternas14} (which are similar to those of \citealt{sheth08}), has been explored in \cite{melvin14}; the most likely reason for this difference is that our $p_{\rm bar}$ threshold of 0.5 tends to identify strong bars, meaning that our work concerns mainly strong bars. The bar fractions used by \cite{cisternas14} are consistent with the total bar fractions of \cite{sheth08}, meaning that their work concerns both strong and weak bars.}, or due to the differences in our matching of AGN and inactive galaxies.  (Our samples are matched in 5 parameters, while \citealt{cisternas14} matched AGN and inactive galaxies in stellar mass and redshift only.)  However, these are only small differences, and the results of \cite{cisternas14} are consistent with our conclusion that bars do not dominate AGN fueling at $z \lesssim 1$. 

However, before ruling out bars as the primary fueling mechanism for supermassive black hole growth, one must ask if the bar--AGN connection can be concealed if a bar dissolves while a black hole is still accreting the bar-funneled gas. In the present analysis, we are assuming that the bar instantaneously funnels gas to the central black hole upon its formation, and moreover, that there is no delay in AGN activity. 

The typical lifetimes of AGN and bars are uncertain. For AGN, the current estimates range from $10^6$ to $10^8$ years (\citealt{haehnelt93, martini04}, see Hanny's Voorwerp in \citealt{keel12} for an example of an AGN with a short lifetime). For bars, early simulations of isolated disc galaxies by \cite{bournaud02} indicate that they are short-lived, with a lifetime of 1--2$\times10^9$ years. The latest simulations of isolated disc galaxies by \cite{athanassoula13b}, however, indicate that bars are long-lived, with a lifetime as long as $10^{10}$ years. This latter result is supported by recent zoom-in cosmological simulations by \cite{kraljic12}, who show that bars formed at $z\approx1$ generally persist down to $z=0$. Despite the uncertainty in both AGN and bar lifetimes, even the shortest bar lifetime is an order of magnitude larger than the longest AGN lifetime, meaning that the bar-AGN connection is not likely to be concealed by short bar lifetimes. 

Small-scale, nuclear bars may also fuel supermassive black hole growth. Unfortunately, we are unable to resolve such small structures in our images. However, work in the local universe shows that nuclear bars are not more frequent in AGN hosts compared to non-AGN hosts \citep{mulchaey97, regan99b, martini99, erwin02, laine02}. This result mirrors that of large-scale bars, suggesting that nuclear bars do not fuel supermassive black hole growth at $z\sim0$ either. Whether nuclear bars can fuel AGN at $z>0$ will be left for future work.  

Interestingly, studies of the relative angle between AGN accretion discs and host galaxy discs are consistent with our interpretation that bars do not fuel AGN. If AGN accretion discs are fueled by bar-funneled gas, then one would expect this gas to have an angular momentum vector that is parallel to the bulk of the gaseous disc of the galaxy. However, it appears that the accretion discs are randomly orientated with respect to their host galaxies \citep{ulvestad84, kinney00, schmitt02, schmitt03, greenhill09}, which fits with our interpretation that bars do not directly fuel AGN.   

This misalignment could be interpreted even more generally---there simply may not be a galactic-scale black hole fueling mechanism. Instead, a collection of processes, including minor mergers \citep[e.g.,][]{kaviraj14a, kaviraj14b}, cooling flows \citep[e.g.,][]{best07} and multi-body interactions with star clusters or clouds \citep[e.g.,][]{genzel94}, may work to transport gas into the vicinity of the black hole. This process, known as ``stochastic fueling,''  \citep{sanders84} has been implemented in models that successfully reproduce observations of low to intermediate luminosity AGN \citep{hopkins06, hopkins13}.

\section{Conclusion}

In this paper, we present a new study on the bar-AGN connection beyond the local universe. We combine {\it Chandra} and Galaxy Zoo: {\it Hubble} data in the AEGIS, COSMOS, and GOODS-S surveys to determine whether AGN are preferentially fed by large-scale bars at $0.2<z<1.0$. 

Using GZH classifications and galaxy structural measurements, we select non-merging, face-on disc galaxies that have sizes large enough to accurately identify large-scale bars. From this face-on disc sample, we identify AGN with {\it X}-ray luminosities $10^{42} ~{\rm erg~s^{-1}} < L_{\rm X} < 10^{44} ~\rm erg~s^{-1}$. We then use a novel multi-parameter technique to construct control samples of non-AGN galaxies robustly matched to the AGN hosts in stellar mass, rest-frame colour, S\'ersic index, effective radius, and redshift. With these samples, we find no statistically significant excess of barred galaxies in AGN hosts (and no excess of AGN in barred galaxies). Specifically, we find that the bar fraction of the AGN sample cannot be greater than twice the bar fraction of the control sample at 99.7\% confidence. The simplest interpretation is that AGN are not preferentially nor directly fed via  {\bf large-scale} bars at $0.2<z<1.0$.

%
%
\section*{Acknowledgements}
%
%

Authors from UC Santa Cruz acknowledge financial support from the NSF Grant AST 08-08133. KS gratefully acknowledges support from Swiss National Science Foundation Grant PP00P2\_138979/1. JT acknowledges support by NASA through {\t Hubble} Fellowship grant HST-HF-51330.01 awarded by the Space Telescope Science Institute, which is operated by the Association of Universities for Research in Astronomy, Inc., for NASA, under contract NAS 5-26555. EA and AB acknowledge financial support to the DAGAL network from the People Programme (Marie Curie Actions) of the European Union's Seventh Framework Programme FP7/2007-2013/ under REA grant agreement number PITN-GA-2011-289313. They also acknowledge financial support from the CNES (Centre National d'Etudes Spatiales - France). RCN acknowledges STFC Rolling Grant ST/I001204/1 "Survey Cosmology and Astrophysics". BDS gratefully acknowledges support from the Oxford Martin School and from the Henry Skynner Junior Research Fellowship at Balliol College, Oxford. LFF and KWW acknowledge support from the UMN GIA program.

EC thanks Ramin A. Skibba, Michael Williams, Sugata Kaviraj, Yicheng Guo, Hassen Yesuf, and Guillermo Barro for useful discussions. The JavaScript Cosmology Calculator \citep{wright06} and TOPCAT \citep{taylor05} were used in the preparation of this paper. We also thank the anonymous referee for a helpful report.

\bibliographystyle{mn2e}

\begin{thebibliography}{90}
\expandafter\ifx\csname natexlab\endcsname\relax\def\natexlab#1{#1}\fi


\bibitem[Aird et al.(2010)]{aird10} Aird, J., Nandra, K., 
Laird, E.~S., et al.\ 2010, \mnras, 401, 2531 

\bibitem[Alonso et 
al.(2013)]{alonso13} Alonso, M.~S., Coldwell, G., \& Lambas, D.~G.\ 2013, \aap, 549, A141 

\bibitem[Athanassoula(1992)]{athanassoula92} 
Athanassoula, E.\ 1992, \mnras, 259, 345 

\bibitem[Athanassoula(2013a)]{athanassoula13a} Athanassoula, E.\ 2013a, 
Secular Evolution of Galaxies, 305 

\bibitem[Athanassoula et al.(2013b)]{athanassoula13b} 
Athanassoula, E., Machado, R.~E.~G., \& Rodionov, S.~A.\ 2013b, \mnras, 429, 1949  

\bibitem[Barnes 
\& Hernquist(1991)]{barnes91} Barnes, J.~E., \& Hernquist, L.~E.\ 1991, \apjl, 370, L65 

\bibitem[Barro et al.(2011)]{barro11a} Barro, G., 
P{\'e}rez-Gonz{\'a}lez, P.~G., Gallego, J., et al.\ 2011, \apjs, 193, 13 

\bibitem[Bauer et al.(2002)]{bauer02} Bauer, F.~E., Alexander, 
D.~M., Brandt, W.~N., et al.\ 2002, \aj, 124, 2351 

\bibitem[Best et al.(2007)]{best07} Best, P.~N., von der 
Linden, A., Kauffmann, G., Heckman, T.~M., 
\& Kaiser, C.~R.\ 2007, \mnras, 379, 894 

\bibitem[Blanton 
\& Roweis(2007)]{blanton07} Blanton, M.~R., \& Roweis, S.\ 2007, \aj, 133, 734 

\bibitem[Bournaud 
\& Combes(2002)]{bournaud02} Bournaud, F., \& Combes, F.\ 2002, \aap, 392, 83 

\bibitem[Brammer et al.(2008)]{brammer08} Brammer, G.~B., van 
Dokkum, P.~G., \& Coppi, P.\ 2008, \apj, 686, 1503 

\bibitem[Bruzual 
\& Charlot(2003)]{Bruzual03} Bruzual, G., \& Charlot, S.\ 2003, \mnras, 344, 1000 

\bibitem[Calzetti et al.(2000)]{calzetti00} Calzetti, D., Armus, 
L., Bohlin, R.~C., et al.\ 2000, \apj, 533, 682 

\bibitem[Cameron(2011)]{cameron11} 
Cameron, E.\ 2011, PASA, 28, 128 

\bibitem[Casteels et al.(2013)]{casteels13} Casteels, K.~R.~V., 
Bamford, S.~P., Skibba, R.~A., et al.\ 2013, \mnras, 429, 1051 

\bibitem[Chabrier(2003)]{chabrier03} Chabrier, G.\ 2003, \pasp, 
115, 763 

\bibitem[Cheung et al.(2012)]{cheung12} 
Cheung, E., Faber, S.~M., Koo, D.~C., et al.\ 2012, \apj, 760, 131

\bibitem[Cheung et al.(2013)]{cheung13} Cheung, E., 
Athanassoula, E., Masters, K.~L., et al.\ 2013, \apj, 779, 162 

\bibitem[Cimatti et al.(2013)]{cimatti13} Cimatti, A., Brusa, M., 
Talia, M., et al.\ 2013, \apjl, 779, LL13 

\bibitem[Cisternas et al.(2011)]{cisternas11} Cisternas, M., 
Jahnke, K., Inskip, K.~J., et al.\ 2011, \apj, 726, 57 

\bibitem[Cisternas et al.(2013)]{cisternas13} Cisternas, M., 
Gadotti, D.~A., Knapen, J.~H., et al.\ 2013, \apj, 776, 50 

\bibitem[Cisternas et al.(2014)]{cisternas14} Cisternas, M., Sheth, 
K., Salvato, M., et al.\ 2014, arXiv:1409.2871 

\bibitem[Civano et al.(2012)]{civano12} Civano, F., Elvis, M., 
Brusa, M., et al.\ 2012, \apjs, 201, 30 

\bibitem[Cooper et al.(2011)]{cooper11} Cooper, M.~C., Aird, 
J.~A., Coil, A.~L., et al.\ 2011, \apjs, 193, 14 

\bibitem[Cooper et al.(2012)]{cooper12} Cooper, M.~C., Griffith, 
R.~L., Newman, J.~A., et al.\ 2012, \mnras, 419, 3018 

\bibitem[Croton et al.(2006)]{croton06} Croton, D.~J., Springel, 
V., White, S.~D.~M., et al.\ 2006, \mnras, 365, 11 

\bibitem[Davis et al.(2003)]{davis03} Davis, M., Faber, S.~M., 
Newman, J., et al.\ 2003, \procspie, 4834, 161 

\bibitem[Davis et al.(2007)]{davis07} Davis, M., Guhathakurta, 
P., Konidaris, N.~P., et al.\ 2007, \apjl, 660, L1 

\bibitem[Di Matteo et al.(2005)]{dimatteo05} Di Matteo, T., 
Springel, V., \& Hernquist, L.\ 2005, \nat, 433, 604 

\bibitem[Dickinson et al.(2003)]{dickinson03} Dickinson, M., 
Giavalisco, M., 
\& GOODS Team 2003, The Mass of Galaxies at Low and High Redshift, 324 

\bibitem[Elvis et al.(2009)]{elvis09} Elvis, M., Civano, F., 
Vignali, C., et al.\ 2009, \apjs, 184, 158 

\bibitem[Erwin 
\& Sparke(2002)]{erwin02} Erwin, P., \& Sparke, L.~S.\ 2002, \aj, 124, 65 

\bibitem[Erwin(2005)]{erwin05} 
Erwin, P.\ 2005, \mnras, 364, 283 

\bibitem[Faber et al.(2003)]{faber03} Faber, S.~M., Phillips, 
A.~C., Kibrick, R.~I., et al.\ 2003, \procspie, 4841, 1657 

\bibitem[Fabian(2012)]{fabian12} Fabian, A.~C.\ 2012, \araa, 50, 455 

\bibitem[Gadotti(2011)]{gadotti11} 
Gadotti, D.~A.\ 2011, \mnras, 415, 3308 

\bibitem[Garmire et al.(2003)]{garmire03} Garmire, G.~P., Bautz, 
M.~W., Ford, P.~G., Nousek, J.~A., 
\& Ricker, G.~R., Jr.\ 2003, \procspie, 4851, 28 

\bibitem[Genzel et al.(1994)]{genzel94} Genzel, R., Hollenbach, 
D., \& Townes, C.~H.\ 1994, Reports on Progress in Physics, 57, 417 

\bibitem[Georgakakis et al.(2006)]{georgakakis06} Georgakakis, A., 
Nandra, K., Laird, E.~S., et al.\ 2006, \mnras, 371, 221 

\bibitem[Giavalisco et al.(2004)]{giavalisco04} Giavalisco, M., 
Ferguson, H.~C., Koekemoer, A.~M., et al.\ 2004, \apjl, 600, L93 

\bibitem[Greenhill et al.(2009)]{greenhill09} Greenhill, L.~J., 
Kondratko, P.~T., Moran, J.~M., \& Tilak, A.\ 2009, \apj, 707, 787 

\bibitem[Griffith et al.(2012)]{griffith12} Griffith, R.~L., 
Cooper, M.~C., Newman, J.~A., et al.\ 2012, \apjs, 200, 9 

\bibitem[Grogin et al.(2011)]{grogin11} Grogin, N.~A., Kocevski, 
D.~D., Faber, S.~M., et al.\ 2011, \apjs, 197, 35 

\bibitem[Guo et al.(2013)]{guo13} Guo, Y., Ferguson, H.~C., 
Giavalisco, M., et al.\ 2013, \apjs, 207, 24 

\bibitem[Haehnelt 
\& Rees(1993)]{haehnelt93} Haehnelt, M.~G., \& Rees, M.~J.\ 1993, \mnras, 263, 168 

\bibitem[Heckman 
\& Best(2014)]{heckman14} Heckman, T., \& Best, P.\ 2014, arXiv:1403.4620 

\bibitem[Ho et al.(1997)]{ho97} 
Ho, L.~C., Filippenko, A.~V., \& Sargent, W.~L.~W.\ 1997, \apj, 487, 591 

\bibitem[Hopkins et al.(2005a)]{hopkins05a} Hopkins, P.~F., 
Hernquist, L., Cox, T.~J., et al.\ 2005a, \apj, 630, 705 

\bibitem[Hopkins et al.(2005b)]{hopkins05b} Hopkins, P.~F., 
Hernquist, L., Martini, P., et al.\ 2005b, \apjl, 625, L71 

\bibitem[Hopkins 
\& Hernquist(2006)]{hopkins06} Hopkins, P.~F., \& Hernquist, L.\ 2006, \apjs, 166, 1 

\bibitem[Hopkins 
\& Quataert(2010)]{hopkins10} Hopkins, P.~F., \& Quataert, E.\ 2010, \mnras, 407, 1529 

\bibitem[Hopkins 
\& Quataert(2011)]{hopkins11} Hopkins, P.~F., \& Quataert, E.\ 2011, \mnras, 415, 1027 

\bibitem[Hopkins et al.(2013)]{hopkins13} Hopkins, P.~F., 
Kocevski, D.~D., \& Bundy, K.\ 2013, arXiv:1309.6321 

\bibitem[Hoyle et al.(2011)]{hoyle11} 
Hoyle, B., Masters, K.~L., Nichol, R.~C., et al.\ 2011, \mnras, 415, 3627 

\bibitem[Huang et al.(2013)]{huang13} Huang, J.-S., Faber, 
S.~M., Willmer, C.~N.~A., et al.\ 2013, \apj, 766, 21 

\bibitem[Hunt 
\& Malkan(1999)]{hunt99} Hunt, L.~K., \& Malkan, M.~A.\ 1999, \apj, 516, 660 

\bibitem[Kartaltepe et al.(2010)]{kartaltepe10} Kartaltepe, J.~S., 
Sanders, D.~B., Le Floc'h, E., et al.\ 2010, \apj, 721, 98 

\bibitem[Kauffmann et al.(2003)]{kauffmann03} Kauffmann, G., 
Heckman, T.~M., Tremonti, C., et al.\ 2003, \mnras, 346, 1055 

\bibitem[Kaviraj(2014a)]{kaviraj14a} Kaviraj, S.\ 2014a, \mnras, 
437, L41 

\bibitem[Kaviraj(2014b)]{kaviraj14b} Kaviraj, S.\ 2014b, 
arXiv:1402.1166 

\bibitem[Keel et al.(2012)]{keel12} Keel, W.~C., Lintott, 
C.~J., Schawinski, K., et al.\ 2012, \aj, 144, 66 

\bibitem[Kinney et al.(2000)]{kinney00} Kinney, A.~L., Schmitt, 
H.~R., Clarke, C.~J., et al.\ 2000, \apj, 537, 152 

\bibitem[Knapen et al.(2000)]{knapen00} Knapen, J.~H., Shlosman, 
I., \& Peletier, R.~F.\ 2000, \apj, 529, 93 

\bibitem[Kocevski et al.(2012)]{kocevski12} Kocevski, D.~D., 
Faber, S.~M., Mozena, M., et al.\ 2012, \apj, 744, 148 

\bibitem[Koekemoer et al.(2007)]{koekemoer07} Koekemoer, A.~M., 
Aussel, H., Calzetti, D., et al.\ 2007, \apjs, 172, 196 

\bibitem[Koekemoer et al.(2011)]{koekemoer11} Koekemoer, A.~M., 
Faber, S.~M., Ferguson, H.~C., et al.\ 2011, \apjs, 197, 36 

\bibitem[Kormendy(1979)]{kormendy79} Kormendy, J.\ 1979, \apj, 
227, 714 

\bibitem[{{Kormendy} \& {Kennicutt}(2004)}]{kk04}
{Kormendy}, J., \& {Kennicutt}, R.~C. 2004, \araa, 42, 603

\bibitem[Kormendy 
\& Ho(2013)]{kormendy13} Kormendy, J., \& Ho, L.~C.\ 2013, \araa, 51, 511 

\bibitem[Koss et al.(2010)]{koss10} Koss, M., Mushotzky, R., 
Veilleux, S., \& Winter, L.\ 2010, \apjl, 716, L125 

\bibitem[Kraljic et al.(2012)]{kraljic12} Kraljic, K., Bournaud, 
F., \& Martig, M.\ 2012, \apj, 757, 60 

\bibitem[Kriek et al.(2009)]{kriek09} Kriek, M., van Dokkum, 
P.~G., Labb{\'e}, I., et al.\ 2009, \apj, 700, 221

\bibitem[Laine et al.(2002)]{laine02} Laine, S., Shlosman, I., 
Knapen, J.~H., \& Peletier, R.~F.\ 2002, \apj, 567, 97 

\bibitem[Laird et al.(2009)]{laird09} Laird, E.~S., Nandra, K., 
Georgakakis, A., et al.\ 2009, \apjs, 180, 102 

\bibitem[Laurikainen et al.(2004)]{laurikainen04} Laurikainen, E., 
Salo, H., \& Buta, R.\ 2004, \apj, 607, 103

\bibitem[Lee et al.(2012a)]{lee12a} 
Lee, G.-H., Park, C., Lee, M.~G., \& Choi, Y.-Y.\ 2012a, \apj, 745, 125 

\bibitem[Lee et al.(2012b)]{lee12b} Lee, G.-H., Woo, J.-H., 
Lee, M.~G., et al.\ 2012b, \apj, 750, 141 

\bibitem[Lilly et al.(2009)]{lilly09} 
Lilly, S.~J., Le Brun, V., Maier, C., et al.\ 2009, \apjs, 184, 218 

\bibitem[Luo et al.(2008)]{luo08} Luo, B., Bauer, F.~E., 
Brandt, W.~N., et al.\ 2008, \apjs, 179, 19 

 \bibitem[Mainieri et al.(2007)]{mainieri07} Mainieri, V., 
Hasinger, G., Cappelluti, N., et al.\ 2007, \apjs, 172, 368 

\bibitem[Malkan et al.(1998)]{malkan98} Malkan, M.~A., Gorjian, 
V., \& Tam, R.\ 1998, \apjs, 117, 25 

\bibitem[Martinet(1995)]{martinet95} Martinet, L.\ 1995, \fcp, 15, 
341 

 \bibitem[Martini 
\& Pogge(1999)]{martini99} Martini, P., \& Pogge, R.~W.\ 1999, \aj, 118, 2646 

\bibitem[Martini et al.(2003)]{martini03} Martini, P., Regan, 
M.~W., Mulchaey, J.~S., \& Pogge, R.~W.\ 2003, \apj, 589, 774 

\bibitem[Martini(2004)]{martini04} Martini, P.\ 2004, Coevolution 
of Black Holes and Galaxies, 169 

\bibitem[{{Masters} {et al.}(2011)}]{masters11} 
{Masters}, K.~L., {Nichol}, R.~C., {Hoyle}, B., et al.\ 2011, \mnras, 411, 2026 

\bibitem[Masters et al.(2012)]{masters12} 
Masters, K.~L., Nichol, R.~C., Haynes, M.~P., et al.\ 2012, \mnras, 424, 2180 

\bibitem[Melvin et al.(2014)]{melvin14} Melvin, T., Masters, K., 
Lintott, C., et al.\ 2014, \mnras, 438, 2882 

\bibitem[Mihos 
\& Hernquist(1996)]{mihos96} Mihos, J.~C., \& Hernquist, L.\ 1996, \apj, 464, 641 

\bibitem[Mulchaey 
\& Regan(1997)]{mulchaey97} Mulchaey, J.~S., \& Regan, M.~W.\ 1997, \apjl, 482, L135 

\bibitem[Muzzin et al.(2013)]{muzzin13} 
Muzzin, A., Marchesini, D., Stefanon, M., et al.\ 2013, \apjs, 206, 8 

\bibitem[Nair \& Abraham(2010b)]{nair10b} 
Nair, P.~B., \& Abraham, R.~G.\ 2010b, \apjl, 714, L260 

\bibitem[Nandra et al.(2005)]{nandra05} Nandra, K., Laird, 
E.~S., Adelberger, K., et al.\ 2005, \mnras, 356, 568 

\bibitem[Nandra et al.(2007)]{nandra07} Nandra, K., Georgakakis, 
A., Willmer, C.~N.~A., et al.\ 2007, \apjl, 660, L11 

\bibitem[Newman et al.(2013)]{newman13} Newman, J.~A., Cooper, 
M.~C., Davis, M., et al.\ 2013, \apjs, 208, 5 

\bibitem[Noguchi(1988)]{noguchi88} Noguchi, M.\ 1988, \aap, 203, 259 

\bibitem[Oh et al.(2012)]{oh12} Oh, S., Oh, K., 
\& Yi, S.~K.\ 2012, \apjs, 198, 4 

\bibitem[Peng et al.(2002)]{peng02} Peng, C.~Y., Ho, L.~C., 
Impey, C.~D., \& Rix, H.-W.\ 2002, \aj, 124, 266

\bibitem[Regan et al.(1995)]{regan95} Regan, M.~W., Vogel, 
S.~N., \& Teuben, P.~J.\ 1995, \apj, 449, 576 

\bibitem[Regan et al.(1999a)]{regan99a} Regan, M.~W., Sheth, K., 
\& Vogel, S.~N.\ 1999, \apj, 526, 97 

\bibitem[Regan \& Mulchaey(1999b)]{regan99b} Regan, M.~W., \& Mulchaey, J.~S.\ 1999, \aj, 117, 2676 

\bibitem[Rix et al.(2004)]{rix04} Rix, H.-W., Barden, M., 
Beckwith, S.~V.~W., et al.\ 2004, \apjs, 152, 163 

\bibitem[Sakamoto et al.(1999)]{sakamoto99} 
Sakamoto, K., Okumura, S.~K., Ishizuki, S., \& Scoville, N.~Z.\ 1999, \apj, 525, 691 

\bibitem[Salpeter(1955)]{salpeter55} Salpeter, E.~E.\ 1955, \apj, 
121, 161 

\bibitem[Sanders(1984)]{sanders84} Sanders, R.~H.\ 1984, \aap, 140, 52 

\bibitem[Sanders et al.(1988)]{sanders88} Sanders, D.~B., Soifer, 
B.~T., Elias, J.~H., et al.\ 1988, \apj, 325, 74 

\bibitem[Schawinski et al.(2009)]{schawinski09} Schawinski, K., 
Virani, S., Simmons, B., et al.\ 2009, \apjl, 692, L19 

\bibitem[Schawinski et al.(2010)]{schawinski10} Schawinski, K., 
Urry, C.~M., Virani, S., et al.\ 2010, \apj, 711, 284 

\bibitem[Schawinski et al.(2011)]{schawinski11} Schawinski, K., 
Treister, E., Urry, C.~M., et al.\ 2011, \apjl, 727, L31 

\bibitem[Schawinski et al.(2012)]{schawinski12} Schawinski, K., 
Simmons, B.~D., Urry, C.~M., Treister, E., 
\& Glikman, E.\ 2012, \mnras, 425, L61 

\bibitem[Schmitt et al.(2002)]{schmitt02} Schmitt, H.~R., 
Pringle, J.~E., Clarke, C.~J., \& Kinney, A.~L.\ 2002, \apj, 575, 150 

\bibitem[Schmitt et al.(2003)]{schmitt03} Schmitt, H.~R., Donley, 
J.~L., Antonucci, R.~R.~J., et al.\ 2003, \apj, 597, 768 

\bibitem[Scoville et al.(2007)]{scoville07} Scoville, N., Aussel, 
H., Brusa, M., et al.\ 2007, \apjs, 172, 1 

\bibitem[Sellwood(2014)]{sellwood14} Sellwood, J.~A.\ 2014, 
Reviews of Modern Physics, 86, 1 

\bibitem[Sheth et al.(2000)]{sheth00} Sheth, K., Regan, M.~W., 
Vogel, S.~N., \& Teuben, P.~J.\ 2000, \apj, 532, 221 

\bibitem[Sheth et al.(2002)]{sheth02} Sheth, K., Vogel, S.~N., 
Regan, M.~W., et al.\ 2002, \aj, 124, 2581 

\bibitem[Sheth et al.(2005)]{sheth05} 
Sheth, K., Vogel, S.~N., Regan, M.~W., Thornley, M.~D., \& Teuben, P.~J.\ 2005, \apj, 632, 217 

\bibitem[Sheth et al.(2008)]{sheth08} Sheth, K., Elmegreen, 
D.~M., Elmegreen, B.~G., et al.\ 2008, \apj, 675, 1141 

\bibitem[Shlosman et al.(1989)]{shlosman89} Shlosman, I., Frank, 
J., \& Begelman, M.~C.\ 1989, \nat, 338, 45 

\bibitem[Shlosman et al.(1990)]{shlosman90} Shlosman, I., 
Begelman, M.~C., \& Frank, J.\ 1990, \nat, 345, 679 

\bibitem[Silverman et al.(2008a)]{silverman08} Silverman, J.~D., 
Green, P.~J., Barkhouse, W.~A., et al.\ 2008a, \apj, 679, 118 

\bibitem[Silverman et al.(2008b)]{silverman08b} Silverman, J.~D., 
Mainieri, V., Lehmer, B.~D., et al.\ 2008b, \apj, 675, 1025 

\bibitem[Simkin et al.(1980)]{simkin80} Simkin, S.~M., Su, 
H.~J., \& Schwarz, M.~P.\ 1980, \apj, 237, 404

\bibitem[Simard et al.(2002)]{simard02} Simard, L., Willmer, 
C.~N.~A., Vogt, N.~P., et al.\ 2002, \apjs, 142, 1 

\bibitem[Simmons et al.(2012)]{simmons12} Simmons, B.~D., Urry, 
C.~M., Schawinski, K., Cardamone, C., \& Glikman, E.\ 2012, \apj, 761, 75 

\bibitem[Simmons et al.(2013)]{simmons13} Simmons, B.~D., 
Lintott, C., Schawinski, K., et al.\ 2013, \mnras, 429, 2199 

\bibitem[Simmons et al.(2014)]{simmons14} Simmons, B.~D., Melvin, 
T., Lintott, C., et al.\ 2014, \mnras, 445, 3466 

\bibitem[Skibba et al.(2012)]{skibba12} Skibba, R.~A., Masters, 
K.~L., Nichol, R.~C., et al.\ 2012, \mnras, 423, 1485 

\bibitem[Springel et al.(2005)]{springel05} Springel, V., Di 
Matteo, T., \& Hernquist, L.\ 2005, \apjl, 620, L79 

\bibitem[Taylor(2005)]{taylor05} Taylor, M.~B.\ 2005, 
Astronomical Data Analysis Software and Systems XIV, 347, 29 

\bibitem[Treister et al.(2012)]{treister12} Treister, E., 
Schawinski, K., Urry, C.~M., \& Simmons, B.~D.\ 2012, \apjl, 758, L39 

\bibitem[Trump et al.(2009)]{trump09} Trump, J.~R., Impey, 
C.~D., Elvis, M., et al.\ 2009, \apj, 696, 1195  

\bibitem[Trump(2013)]{trump13} 
Trump, J. R. 2013, Proc. from Galaxy Mergers in an Evolving Universe, Ed. by W.-H. Sun, C. K. Xu, N. Z. Scoville, \& D. B. Sanders, ASPC, 477, 227

\bibitem[Ueda et al.(2003)]{ueda03} Ueda, Y., Akiyama, M., 
Ohta, K., \& Miyaji, T.\ 2003, \apj, 598, 886 

\bibitem[Ulvestad 
\& Wilson(1984)]{ulvestad84} Ulvestad, J.~S., \& Wilson, A.~S.\ 1984, \apj, 285, 439 

\bibitem[Wada 
\& Habe(1992)]{wada92} Wada, K., \& Habe, A.\ 1992, \mnras, 258, 82 

\bibitem[Willett et al.(2013)]{willett13} 
Willett, K.~W., Lintott, C.~J., Bamford, S.~P., et al.\ 2013, \mnras, 435, 2835 

\bibitem[Williams et al.(2014)]{williams14} Williams, C.~C., 
Giavalisco, M., Cassata, P., et al.\ 2014, \apj, 780, 1 

\bibitem[Wright(2006)]{wright06} Wright, E.~L.\ 2006, \pasp, 
118, 1711 

\bibitem[Xue et al.(2011)]{xue11} Xue, Y.~Q., Luo, B., 
Brandt, W.~N., et al.\ 2011, \apjs, 195, 10 

\bibitem[Zurita et 
al.(2004)]{zurita04} Zurita, A., Rela{\~n}o, M., Beckman, J.~E., \& Knapen, J.~H.\ 2004, \aap, 413, 73 

\end{thebibliography}

\end{document}